\documentclass[floats,prd,twocolumn,showpacs,amssymb]{revtex4}
\usepackage{amssymb}
\usepackage{graphics}
\usepackage{amsmath}
\usepackage{amsfonts}
\usepackage{bm}

\begin{document}
\newcommand*{\be}{\begin{equation}}
\newcommand*{\ee}{\end{equation}}

\title{Primordial magnetic fields and gravitational baryogenesis in
nonlinear electrodynamics}

\author{Herman J. Mosquera Cuesta$^{1,2}$ and Gaetano Lambiase$^{3,4}$}

\affiliation{$^1$\mbox{Instituto de Cosmologia, Relatividade e
Astrof\'{\i}sica (ICRA-BR), Centro Brasileiro de Pesquisas
F\'{\i}sicas (CBPF)}\\ \mbox{Rua Dr. Xavier Sigaud 150, CEP
22290-180, Rio de Janeiro, RJ, Brazil } \\
$^2$\mbox{ICRANet International Coordinating Centre, P.le della 
Repubblica 10, 65100, Pescara, Italy }\\
$^3$\mbox{Dipartimento di Fisica "E.R. Caianiello", Universit\'a
di Salerno, 84081 Baronissi (Sa), Italy.} \\
$^4$INFN, Sezione di Napoli, Italy.}
\date{\today}
\renewcommand{\theequation}{\thesection.\arabic{equation}}
\begin{abstract}
The amplification of the primordial magnetic fields and the gravitational baryogenesis, a
mechanism that allows to generate the baryon asymmetry in the
Universe by means of the coupling between the Ricci scalar
curvature and the baryon current, are reviewed in the framework of
the nonlinear electrodynamics.
To study the amplification of the primordial
magnetic field strength, we write down the gauge invariant wave equations and then
solve them (in the long wavelength
approximation) for three different eras of the Universe: de
Sitter, the reheating and the radiation dominated era.

Constraints on parameters entering the nonlinear electrodynamics are obtained
by using the amplitude of the observed galactic magnetic fields and the baryon
asymmetry, which are characterized by the dimensionless parameters $r\sim
10^{-37}$ and $\eta_B\lesssim 9\times 10^{-11}$, respectively.
\end{abstract}
 \vskip -1.0 truecm
 \maketitle


\section{Introduction}
\setcounter{equation}{0}

With the aim to build up a classically singularity-free theory of
the electron, that is a theory in which infinite physical
quantities are avoided, Born and Infeld \cite{bohr} proposed a
model in which additional terms or modifications of the standard
electrodynamics were included. To prevent the infinite self energy
of point particles (as follows from standard electrodynamics),
they introduced an upper limit on the electric field strength and
considered the electron as an electric particle with finite
radius. In successive investigations, other examples of nonlinear
electrodynamics Lagrangians were proposed by Plebanski, who also
showed that Born-Infeld model satisfy physically acceptable
requirements \cite{plebanski}. Consequences of nonlinear
electrodynamics have been studied in many contexts, such a, for
example, cosmological models \cite{cosmology}, black holes and
wormhole physics \cite{BH,wormhole}, and astrophysics
\cite{applications}.

Recently, the nonlinear electrodynamics has been also invoked as
an available framework for generating the primordial magnetic
fields in the Universe \cite{kunze,cea}. The latter, indeed, is a
still open problem of the modern cosmology, and although many
mechanisms have been proposed, this issue is far to be solved.
Seed of magnetic fields may arise in different contexts, e.g.
cosmological phase transitions of the early Universe
\cite{vachaspati}, string cosmology \cite{gasperini}, inflationary
models of the Universe \cite{ratra,turner}, nonminimal
electromagnetic-gravitational coupling
\cite{opher,bamba,prasanna}, gauge invariance breakdown
\cite{turner,mazzitelli,prasanna}, density perturbations
\cite{riotto}, gravitational waves in the early Universe
\cite{tsagas}, lorentz violation \cite{bertolami}, cosmological
defects \cite{vilenkin}, electroweak anomaly \cite{joyce},
temporary electric charge nonconservation \cite{silk}, trace
anomaly \cite{dolgovPRD}, parity violation of the weak
interactions \cite{semikoz}, Biermann type battery seed effect
\cite{biermann}. Once these seeds are generated, they must be
amplified by means of some mechanism. Promising candidates are the
dynamo mechanism \cite{zeldovich,Parker} and the protogalactic
collapse and differential rotation \cite{piddington}. The first
mechanism allows an enhancement of the (preexisting) magnetic
strength from $\sim 10^{-20}$ G to $\sim 10^{-6}-10^{-5}$G, the
present (observed in galaxies and galaxy clusters) strength, the
second one instead allows an amplification from $\sim 10^{-10}$ G
to $\sim 10^{-6}-10^{-5}$G. For a review, see \cite{grasso}.
Moreover, the presence of magnetic fields in the Universe has
important cosmological consequences as, for example, the
generation of anisotropies in CMB \cite{barrow}, and the
primordial abundances of the light elements [Big Bang
Nucleosynthesis (BBN)] \cite{greestein}.

In this paper, besides to study the amplification of primordial magnetic fields
in the context of nonlinear electrodynamics, we also discuss the possibility
that nonlinear electrodynamics might provide a framework for the so-called {\it gravitational}
baryogenesis. The latter is related to the origin of the baryon
number asymmetry, which is, as well known, a still open problem of
the particle physics and cosmology \cite{kolb}. BBN \cite{burles}
and measurements of CMB combined with the large structure of the
Universe \cite{bennet} indicate that the order of magnitude of such an
asymmetry is
 \[
\eta_B \equiv \frac{n_B-n_{\bar B}}{s}\lesssim 9\,\, 10^{-11}\,,
 \]
where $n_B$ ($n_{\bar B}$) is the baryon (antibaryon) number
density, and $s$ the entropy of the Universe. Conventionally, the necessary requirements
for a (CPT invariant) theory able to generate the baryon asymmetry are dictated by Sakharov's conditions
\cite{sakharov}: 1) there must exist processes that violate the baryon number; 2) the discrete symmetries
C and CP must be violated; 3) departure from thermal equilibrium.
However, none of the Sakharov's
conditions is obligatory \cite{dolgov}.
In fact, as shown in \cite{cohen}, a dynamical
violation of CPT (which implies a different spectrum of particles and antiparticles)
may give rise to the baryon number asymmetry also
in a regime of thermal equilibrium.

The paper is organized as follows. In next Section we study the
amplification of the primordial magnetic fields in the framework of
the nonlinear electrodynamics. We shall investigate the case in
which the Lagrangian is of the form $L\sim X+\gamma X^\delta$,
where $X=F_{\mu\nu}F^{\mu\nu}/4$, and $\gamma$ and $\delta$ are
free parameters. In Section III, after a short review of the
gravitational baryogenesis mechanism \cite{steinhardt}, we
investigate the possibility to generate, during the radiation
dominated era, the observed baryon asymmetry if effects of
nonlinear electrodynamics are taking into account. Section IV is devoted
to the analysis of the amplification of primordial magnetic fields and of the gravitational baryogenesis
for the nonlinear electrodynamics whose Lagrangian is
of the form ${\cal L}\sim  X+\gamma/X$. Conclusions are shortly
discussed in Section V.

\section{Field Equations in nonlinear Electrodynamics and Primordial
Magnetic Field}
 \setcounter{equation}{0}

In this Section we shall study the amplification of the primordial
magnetic field for the case in which the electromagnetic field is
described by a nonlinear electrodynamics. The Lagrangian density we consider
is \cite{notaY}
\begin{equation}\label{Lagr}
 L(X)=-C X-\gamma X^\delta\,.
\end{equation}
where $\gamma$ and $\delta$ are free parameters that with
appropriate choice reproduce the well know Lagrangian already
studied in literature. $\gamma$ has dimensions
[energy]$^{4(1-\delta)}$. The case $C=1$ and $\gamma=0$
corresponds to the standard linear electrodynamics. The primordial
magnetic field in nonlinear electrodynamics has been studied
recently by Kunze \cite{kunze} and Campanelli et al. \cite{cea}.
Their study refer mainly during the Inflationary era of the
Universe evolution. Our approach follows the paper \cite{turner},
in which the electromagnetic field evolution is analyzed during
the de Sitter, Rehating and radiation dominated eras. Moreover, we
derive a wave equation for the electromagnetic field strength
tensor $F_{\mu\nu}$.

In the seminal paper by Turner and Widrow (TW) \cite{turner}, it
was suggested that a magnetic field might be generated by quantum
fluctuations during an inflationary epoch, and it could be
sustained after the wave length of interest crossed beyond the
horizon giving the observed field today \cite{turner}. This model
invokes a coupling among the electromagnetic field and the scalar
($R$) and (Ricci and Riemann) tensor curvatures, which break the
conformal invariance. According to TW paper, since the Universe is
a good conductor (during its evolution), one expects that the
magnetic flux is preserved even if the primordial magnetic field
evolves. This physical behavior suggests the definition of the
parameter $r=\rho_B/\rho_\gamma$, which remains (with good
approximation) constant and provides an invariant measure of the
magnetic field strength. Here $\rho_B$ is the energy density of
the magnetic field, and $\rho_\gamma=\pi^2T^4/25$ is the energy
density of the cosmic microwave background radiation. In order to
explain the present value of $r\approx 1$ for galaxies, one needs
a pre-galactic magnetic field which corresponds to $r\simeq 10^{-37}$
if dynamo amplifications are invoked, and $r\simeq 10^{-8}$ if the galactic
magnetic fields are generated, in the collapse of the
protogalactic cloud, by means of the compression of the primordial
magnetic field. In the last case, the dynamo processes are not
necessary (see, for example, Refs. \cite{turner,kunze}).

The action we consider is the electromagnetic field minimally
coupled to gravity
 \begin{equation}\label{actionL}
S= \int d^4x\sqrt{-g}\left(\frac{R}{16 \pi G}
    +\frac{1}{4\pi} L(X) \right)\,,
 \end{equation}
The nonlinearity breaks the conformal invariance, which is the necessary condition
for amplifying the primordial magnetic fields (in fact, the minimal
coupling of electromagnetic fields to a four-dimensional
background is invariant under conformal transformations of the
metric; therefore, the time evolution of the conformally flat
metric, as the Friedamn-Robertson-Walker metric, does not affect
the electromagnetic fluctuations, and no amplifications occur).


The field equations for the electromagnetic fields are
\begin{equation}\label{fieldequationL}
  \nabla_\rho F^{\rho\sigma}=-\frac{\nabla_\mu L_{X}}{L_X}  F^{\mu \sigma}\,,
\end{equation}
\begin{equation}\label{bianchi}
  \nabla_\mu F_{\nu\lambda}+\nabla_\nu F_{\lambda\mu}+\nabla_\lambda
  F_{\mu\nu}=0\,.
\end{equation}

Eq. (\ref{bianchi}) are the Bianchi identities and $L_X=dL/dX$.
The wave equation for $F_{\mu\nu}$ follows by applying
$\nabla_\lambda$ to Eq. (\ref{bianchi}) and then using Eq.
(\ref{fieldequationL}). One gets

\begin{eqnarray} \label{5}
  \Box F_{\nu\lambda}+[\nabla^\mu, \nabla_\nu]F_{\lambda\mu}
  -[\nabla^\mu, \nabla_\lambda]F_{\nu\mu} &=& \\
  -\nabla_\nu \left( \frac{\nabla_\alpha L_{X}}{L_X}F^{\alpha}_{\,\,\,
  \lambda}\right)+(\nu\leftrightarrow \lambda)\,, \nonumber
\end{eqnarray}
where $\Box =\nabla^\mu\nabla_\mu$ and $[.,.]$ is the commutator.

Using 
\begin{itemize}
 \item 1) the cyclic identities of the Riemann tensor $
R_{\rho\alpha\beta\gamma}+R_{\rho\gamma\alpha\beta}+R_{\rho\beta\gamma\alpha}=0$,
\item 2) the Ricci identity $[\nabla^\mu, \nabla_\nu]F_{\alpha\mu}=
R_{\rho\alpha\nu\mu}F^{\rho\mu}+R^\rho_{\phantom{\rho}\nu}
F_{\alpha\rho}$, 
\item 3) the fact that the Riemann tensor
can be written in terms of the Ricci tensor and the scalar
curvature $R$ as (this is true because in a system of coordinates
in which the metric is conformal to the Minkowski one, the Weyl
tensor $C_{\lambda\mu\nu\rho}$ vanishes \cite{weinberg})
\end{itemize}

  \begin{eqnarray}
 R_{\lambda\mu\nu\rho}&=&\frac{1}{2}(g_{\lambda\nu}R_{\mu\rho}
- g_{\lambda\rho}R_{\mu\nu}
 -g_{\mu\nu}R_{\lambda\rho}+g_{\mu\rho}R_{\lambda\nu}) \nonumber \\
  & & - \frac{R}{6}(g_{\lambda\nu}g_{\mu\rho}-
 g_{\lambda\rho}g_{\mu\nu})\,, \nonumber
 \end{eqnarray}
one can rewrite Eq. (\ref{5}) as
 \begin{eqnarray}
  \Box F_{\alpha\lambda}-\frac{R}{3}F_{\alpha\lambda}&=&\frac{1}{a^2}
\Box_\eta  F_{\alpha\lambda}= \label{eqGaugeInv} \\
  &=& -\nabla_\alpha \left( \frac{\nabla_\mu L_{X}}{L_X}F^{\mu}_{\,\,\,
  \lambda}\right)+(\alpha \leftrightarrow \lambda)\,,\nonumber
\end{eqnarray}
where $\Box_\eta=\eta^{\mu\nu}\partial_\mu\partial_\nu$ is the
D'Alambertian in the Minkowski spacetime. Eq. (\ref{eqGaugeInv})
is gauge invariant.

We work in the conformal Friedman-Lema\^itre-Robertson-Walker 
(FLRW) metric
\begin{equation}\label{conf-metric}
  g_{\mu\nu}=a^2(\eta) \, \mbox{diag} (1, -1, -1, -1)\,,
\end{equation}
where $a(\eta)$ is the scale factor. The field strength tensor
$F_{\mu\nu}$ in a curved spacetime has components
\begin{equation}\label{Fmunu}
  F_{\mu\nu}=a^2(\eta) \begin{pmatrix}
                       0  & -E_x & -E_y & -E_z  \\
                       E_x & 0 & B_z & -B_y \\
                       E_y & -B_z & 0 & B_x \\
                       E_z & B_y & -B_x & 0 \\
                       \end{pmatrix} \,.
\end{equation}
We shall set $E_i=0$.

Eqs. (\ref{eqGaugeInv}) are very involved. To evaluate the
magnetic field strength during the three eras we are interested
in, i.e. {\it de Sitter (dS)}, {\it Reheating (RH)} and {\it
Radiation Dominated (RD)} phases of the Universe, we concern with
the evolution of the magnetic field fluctuations whose wavelengths
are well outside the horizon, $L_{phys}=aL\gg H^{-1}$ or $k\eta\ll
1$ \cite{turner}. In this approximation, all spatial derivatives
will be neglected (long wave-length approximation). Moreover, we
shall assume that the direction of the magnetic field is fixed.
Therefore, using the relations $F_{ij}=\varepsilon_{ijk}(a^2
B_k)$, and the notation $ |{\bf F}(\eta)|\equiv F = a^2(\eta)|{\bf
B}(\eta)|$, the field equation (\ref{eqGaugeInv}) reduces to the
form

\begin{equation}\label{EqWaveF}
  \left[C+\gamma\delta\left(\frac{F^2}{2a^4}\right)^{\delta-1}\right]
  F''+
\end{equation}
 \[
   +\gamma\delta(\delta-1)\left(\frac{F^2}{2a^4}\right)^{\delta-2}
  \left(\frac{F^2}{2a^4}\right)^\prime {\cal H} F=0 \,.
 \]

The prime means derivative with respect to the conformal time
$\eta$ and ${\cal H}=a'/a$ is the Hubble parameter.

It turns out convenient to express Eq. (\ref{EqWaveF}) in terms of the scale
factor $a$. Since the scale factor varies as $a(\eta)=a_{(\alpha)} \eta^\alpha$, where $\alpha=-1, +2, +1$ during the $dS$, $RH$ and $RD$ eras, respectively, while the constants
$a_{(\alpha)}$, that are different for three eras, are explicitly specified in (\ref{ScaleFactorInfl}), (\ref{ScaleFactRH}), and (\ref{ScaleFactRD}), we get
\begin{equation}\label{EqWaveFa}
  \frac{d^2
  F}{da^2}+\left[1-\frac{1}{\alpha}+4(\delta-1){\cal F}\right]\frac{1}{a}
  \frac{dF}{da}-8(\delta-1)\frac{\cal F}{a^2}F=0\,,
\end{equation}
where
 \[
 {\cal F}\equiv \frac{\gamma\delta
 \displaystyle{\left(\frac{F^2}{2a^4}\right)^{\delta-1}}}
 {C+\gamma\delta\displaystyle{\left(\frac{F^2}{2a^4}\right)^{\delta-1}}}\,.
 \]
The complex structure of the differential equation
(\ref{EqWaveFa}) does not allow to determine
exact solutions. We therefore assume that during the $dS$, $RH$
and $RD$ eras the $(F^2/a^4)$-term is dominant, which means ${\cal
F}\sim 1$. In this regime, a solution of (\ref{EqWaveFa}) is of
the form
\begin{equation}\label{solution}
  F(\eta) = F_{(\beta)} a^\beta\,,
\end{equation}
where $F_{(\beta)}$, which is a constant, and $\beta$ assume different values for
each different phase of the Universe evolution.

\subsection{Inflationary de Sitter (dS) phase ($\alpha=-1$)}

The scale factor for this epoch of the Universe is

\begin{equation}\label{ScaleFactorInfl}
  a(\eta)=- a_{dS} \, \eta^{-1}\sim -\frac{1}{H_{dS}\eta}\,,
\end{equation}
where $H_{dS}\sim 3\times 10^{24}$eV. Eq. (\ref{solution}) reads
\begin{equation}\label{F(a)Infl}
  F \sim a^{\beta_{dS}}\,.
\end{equation}
The exponent $\beta_{dS}$ is given by
\begin{equation}\label{beta-dS}
  \beta_{dS}\equiv p_{\pm}=\frac{3}{2}-2\delta\pm
  \sqrt{4\delta^2+2\delta-\frac{23}{4}}\,.
\end{equation}

\subsection{Reheating (RH) phase ($\alpha=2$)}

The scale factor for this stage of the Universe is given by \cite{bertolami}
\begin{equation}\label{ScaleFactRH}
  a(\eta)=a_{RH}\, \eta^2 \sim \frac{1}{4}H_0^2R_0^3\, \eta^2\,,
\end{equation}
where $R_0\sim 10^{26}h_0^{-1}$m is the present Hubble radius of
the Universe, and $H_0\sim 100 h_0 $km/Mpc sec is the Hubble
parameter today. The solution (\ref{solution}) is of the form
\begin{equation}\label{F(a)RH}
 F \sim  a^{\beta_{RH}}\,,
\end{equation}
where
\begin{equation}\label{beta-RH}
  \beta_{RH}\equiv q_{\pm}=\frac{9}{4}-2\delta\pm
  \sqrt{4\delta^2-\delta-\frac{47}{16}}\,.
\end{equation}

\subsection{Radiation Dominated (RD) phase ($\alpha=1$)}

In this last case, the scale factor of the Universe is

\begin{equation}\label{ScaleFactRD}
  a= a_{RD} \, \eta \sim H_0R_0^2\, \eta\,.
\end{equation}
The solution for $F$ is
\begin{equation}\label{F(a)RD}
  F \sim  a^{\beta_{RD}}\,,
\end{equation}
where
\begin{equation}\label{beta-RD}
  \beta_{RD}=\frac{3}{2}-2\delta\pm
  \sqrt{4\delta^2-2\delta-\frac{7}{4}}\,.
\end{equation}

The solutions (\ref{F(a)Infl}), (\ref{F(a)RH}), and (\ref{F(a)RD}) have been determined for ${\cal F}\sim 1$. By using (\ref{solution}) one infers that the regime we concern
applies for amplitudes of the magnetic field such that
 \[
 |{\bf B}(\eta)| \gg B_0\,, \qquad B_0 \equiv
 \sqrt{2}\left(\frac{C}{\gamma |\delta|}\right)^{\frac{1}{2(\delta-1)}}\,,
 \]
or equivalently, in terms of the conformal time, it applies for conformal time $\eta$ larger than
$\eta_*$,
 \[
 \eta \gg \eta_*\,, \quad
\eta_*\equiv \displaystyle{\frac{1}{a_{(\alpha)}}\left[\frac{\sqrt{2}}{F_{(\beta)}}
\left(\frac{C}{\gamma |\delta|}\right)^{1/2(\delta-1)}\right]^{1/(\alpha(\beta-2))} }\,,
 \]
where $a_{(\alpha)}=a_{dS}, a_{RH}, a_{RD}$ are defined in Eqs. (\ref{ScaleFactorInfl}), (\ref{ScaleFactRH}), and (\ref{ScaleFactRD}), and $\beta=\beta_{dS}, \beta_{RH}, \beta_{RD}$
are given by Eqs. (\ref{beta-dS}), (\ref{beta-RH}) and (\ref{beta-RD}).

\vspace{0.2in}

The above solutions for $F=F_k(a)$ allow to estimate
the strength of the primordial magnetic field. According to
Turner-Widrow model \cite{turner}, if one assumes that the
Universe had gone through a period of inflation at GUT scale
($M_{GUT}\sim 10^{16}\div 10^{17}$GeV) and that fluctuations of
the electromagnetic field have come out from the horizon where the
Universe had gone through about 55 e-folding of inflation, then
\cite{turner}
\begin{eqnarray}\label{r}
  r &\approx & (7\times
  10^{25})^{-2(p+2)}\left(\frac{M_{GUT}}{m_{Pl}}\right)^{4(q-p)/3}\times \\
  &\times& \left(\frac{T_{RH}}{m_{Pl}}\right)^{2(2q-p)/3}
  \left(\frac{T_{*}}{m_{Pl}}\right)^{-8q/3}\lambda^{-2(p+2)}_{Mpc}\,,
  \nonumber
\end{eqnarray}
where $T_{RH}$ is the reheating temperature, $T_*$ is the
temperature at which plasma effects become dominant (i.e. the
Universe first becomes a good conductor), and $m_{Pl}\sim
10^{19}$GeV is the Planck mass. Finally, $p=p_\pm$ and $q=q_\pm$
are the exponents of the scale factor $a(\eta)$ during the dS and
RH epochs (see Eqs. (\ref{F(a)Infl}) and (\ref{F(a)RH})).
Results are independent on the parameter $\gamma$.

The temperature $T_*$ can be estimated via reheating processes
\cite{turner} $T_*=min\{(T_{RH}M_{GUT})^{1/2};
(T^2_{RH}m_{Pl})^{1/3}\}$, and for $T<T_*$ $\rho_B$ evolves as
$\rho_B\sim a^{-4}$. Notice that the reheating
temperature $T_{RH}$ is given by $T_{RH}=\{10^{9}\mbox{GeV},
M_{GUT}\}$ \cite{turner}. Imposing that $r\sim 10^{-37}$, we infer the values
for the parameter $\delta$ yielding the observed strength of the cosmological
magnetic field. Results are reported in Tables I and II.

\begin{table}
\caption{Values of $\delta$ for $r\sim 10^{-37}$ at 1Mpc and for
$M_{GUT}\sim 10^{17}$GeV and $T_{RH}\sim 10^{15}- 10^{17}$GeV. The
cases $p_+, q_+$ and $p_+, q_-$ do not admit solutions.}
\begin{ruledtabular}
\begin{tabular}{lllll}
 $p, q$     & $M_{GUT}$(GeV) & $T_{RH}$(GeV) & $T_*$(GeV) & $\delta\sim $
\vspace{0.09in}  \\ \hline\hline
 $p_-, q_+$ &  $10^{17}$ & $10^{15}$ & $10^{15}$ &  $1.280$ \\
            &            & $10^{16}$ & $10^{16}$ & 1.278 \\
            &            & $10^{17}$ & $10^{16}$ & 1.265
            \vspace{0.09in}\\ \hline
 $p_-, q_-$ & $10^{17}$  & $10^{15}$ & $10^{15}$ & 1.315  \\
            &            & $10^{16}$ & $10^{16}$ & 1.295 \\
            &            & $10^{17}$ & $10^{16}$ & 1.297\\
\end{tabular}
\end{ruledtabular}
\end{table}

\begin{table}
\caption{Values of $\delta$ for $r\sim 10^{-37}$ at 1Mpc and for
$M_{GUT}\sim 10^{16}, 10^{17}$GeV and $T_{RH}\sim 10^{9}$GeV,
$T_*\sim 10^{12}$GeV. The cases $p_+, q_+$ and $p_+, q_-$ do not
admit solutions.}
\begin{ruledtabular}
\begin{tabular}{lllll}
 $p, q$     & $M_{GUT}$(GeV) & $T_{RH}$(GeV) & $T_*$(GeV) & $\delta\sim $
 \vspace{0.09in}
 \\ \hline\hline
 $p_-, q_+$ &  $10^{17}$ & $10^{9}$ & $10^{12}$ & 1.331 \\
            &  $10^{16}$ &          &           & 1.360 \\  \hline
 $p_-, q_-$ &  $10^{17}$ & $10^9$   & $10^{12}$ & 1.375  \\
            &  $10^{16}$ &          &           & 1.382 \\
\end{tabular}
\end{ruledtabular}
\end{table}

\begin{figure}
\resizebox{6cm}{!}{\includegraphics{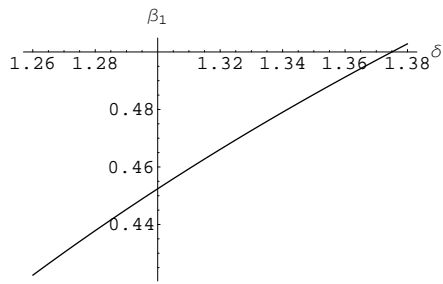}} \caption{The plot
represents $\beta_1\equiv \beta_{RD}^{(+)}=\frac{3}{2}-2\delta+
\sqrt{4\delta^2-2\delta-\frac{7}{4}}$ vs $\delta$.}
 \label{beta1}
\end{figure}

\begin{figure}
\resizebox{6cm}{!}{\includegraphics{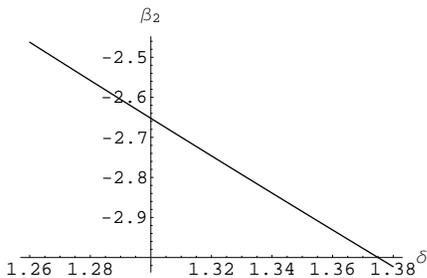}} \caption{The plot
represents $\beta_2\equiv \beta_{RD}^{(-)}=\frac{3}{2}-2\delta-
\sqrt{4\delta^2-2\delta-\frac{7}{4}}$ vs $\delta$.}
 \label{beta2}
\end{figure}

Some comments are in order. First, during the radiation dominated
era, the plasma effects induce a rapid decay of the electric
field, whereas the magnetic field remains \cite{turner}. Moreover,
the functions $F(\eta)$ have been obtained for a cosmological
background which evolves according to standard Cosmology. In
particular, during the radiation dominated era the scale factor
evolves according to the power law $a\sim \eta \sim t^{1/2}$ ($t$
is the cosmic time). The "magnetic" component of the energy
density, therefore, is assumed negligible with respect to the
radiation energy density: $\rho_{total}=\rho_{rad}+\rho_B\simeq
\rho_{rad}(=\frac{\pi^2 g_*}{30}T^4)$. The validity of the
condition $\rho_B < \rho_{rad}$, that will be discussed in the
next Section when we will study the origin of baryon asymmetry,
yields a constraint on the temperature at which the nonlinear
effects are active. Yet, in order that predictions of the standard
Cosmology (such as BBN, CMB, and large scale structure formation)
remain unaltered, we assume that after the conformal time ${\tilde \eta}$
(or after the cosmic time  ${\tilde t}$ or the temperature
${\tilde T}$) corrections to the standard linear electrodynamics
vanish, i.e. $\gamma=0$ for $\eta> {\tilde \eta}$, and $\gamma\neq
0$ for $\eta_{RD}< \eta < {\tilde \eta}$, where $\eta_{RD}$ is the
time when radiation-dominated era starts (that, in our model, it
does coincide with the end of reheating). Of course, ${\tilde
\eta} \ll \eta_{end}$, where $\eta_{end}$ corresponds to the end
of the radiation dominated era. 
Figs. \ref{beta1} and \ref{beta2}
show the behaviors of $\beta_{RD}^{(\pm)}$ for the range of values for $\delta$
reported in Tables I and II.


\section{Gravitational Baryogenesis}

To begin with, we shortly recall the main topics of the
gravitational baryogenesis. The latter, as already pointed out, is a mechanism for
generating the baryon number asymmetry during the expansion of the
Universe by means of a dynamical breaking of CPT (and CP)
\cite{steinhardt}. In this approach the thermal equilibrium is
preserved. The interaction responsible for CPT violation is given
by a coupling between the derivative of the Ricci scalar curvature
$R$ and the baryon current $J^\mu$ \cite{noteJ}

\begin{equation}\label{riccicoupling}
  \frac{1}{M_*^2}\int d^4x \sqrt{-g}J^\mu\partial_\mu R\,,
\end{equation}
where $M_*$ is the cutoff scale characterizing the effective
theory. If there exist interactions that violate the baryon number
$B$ in thermal equilibrium, then a net baryon asymmetry can be
generated and gets frozen-in below the decoupling temperature
$T_D$.

From (\ref{riccicoupling}) it follows
 \[
M_*^{-2} (\partial_\mu R)J^\mu=M_*^{-2} {\dot R}(n_B-n_{\bar B})\,,
 \]
where ${\dot R}=dR/dt$. Therefore the effective chemical potential for
baryons and antibaryons is $\mu_B={\dot R}/M_*^2 =-\mu_{\bar B}$,
and the net baryon number density at the equilibrium turns out to
be (as $T\gg m_B$, where $m_B$ is the baryon mass) $n_B=g_b\mu_B
T^2/6$. $g_b\sim {\cal O}(1)$ is the number of intrinsic degrees
of freedom of baryons. The baryon number to entropy ratio, that defines
the baryon asymmetry, is therefore
\cite{steinhardt}
\begin{equation}\label{nB/s}
 \eta_B= \frac{n_B}{s}\simeq -\frac{15g_b}{4\pi^2g_*}\frac{\dot R}{M_*^2
  T}\Big|_{T_D}\,,
\end{equation}
where $s=2\pi^2g_{*s}T^3/45$, and $g_{*s}$ counts the total
degrees of freedom for particles that contribute to the entropy of
the Universe. $g_{*s}$ takes values very close to the total
degrees of freedom of effective massless particles $g_*$, i.e.
$g_{*s}\simeq g_*\sim 106$. $\eta_B$ does not vanish
provided that the time derivative of the Ricci scalar is non
vanishing.

In the context of General Relativity, the Ricci scalar
and the trace $T_g$ of the energy-momentum tensor ($T_g^{\mu\nu}$)
are related by the relation
 \[
R=-8\pi GT_g =-8\pi G(1-3w)\rho\,,
 \]
where $\rho$ is the matter density, $w=p/\rho$
is the adiabatic parameter, $p$ the pressure, and $T_g=T_{g\, \mu}^\mu$. ${\dot R}$
is zero in the radiation dominated epoch of the standard FLRW
cosmology, because (in the limit of {\it exact} conformal
invariance) $w=1/3$. However, deviations from the standard
electrodynamics prevent the Ricci curvature and its first time
derivative to vanish (as seen from the point of view of the new
structure of the energy-momentum tensor). Therefore a net baryon
asymmetry may be generated also during the radiation dominated era
(for other applications and scenarios see
\cite{steinhardt,li,scalar,lambiase}).

\subsection{Gravitational Baryogenesis in nonlinear Electrodynamics}

We wish now discuss the origin of the baryon asymmetry in the framework of
the nonlinear electrodynamics. The epoch of the Universe we refer to is the
radiation dominated era. As pointed out at the end of Sect. II, we
assume that from the beginning of the radiation dominated era to
time ${\tilde t}$ the nonlinear terms of electromagnetism are non
zero. The latter may break the conformal invariance and therefore
$1-3w \neq 0$, or equivalently, the trace of the energy momentum
tensor does not vanishes. As a consequence, $R$ and ${\dot R}$ are
different from zero. In fact, by making use of the expression for
the energy-momentum tensor
\begin{equation}\label{LXenemom}
  T_{g\,\,\mu\nu}=\frac{1}{4\pi}\left[
  \frac{\partial L}{\partial
  X}F_{\mu\alpha}F^{\alpha}_{\,\,\,
  \nu}+g_{\mu\nu}L \right]\,,
\end{equation}
we infer that the trace $T_g$ is given by
 \[
 T_g=-\frac{\gamma(\delta-1)}{\pi}X^\delta\,.
 \]

Eq. (\ref{F(a)RD}) implies that ${\dot X}=(\beta_{RD}-2)H B^2$,
where $H={\dot a}/a$.

By making use of the Einstein field equations
 \begin{equation}\label{einstein}
H=\displaystyle{\frac{\pi}{3m_P}\sqrt{\frac{4\pi g_*}{5}}T^2}\,,
 \end{equation}
the parameter $\eta_B$ characterizing the baryon
asymmetry (see Eq. (\ref{nB/s})) can be cast in the form
\begin{equation}\label{nB/s-1}
  \eta_B = 8g_b\sqrt{\frac{5}{\pi g_*}}
  (\beta_{RD}-2)\delta(\delta-1)\gamma
  \left(\frac{B^2}{2}\right)^\delta \frac{T_D}{M_*^2 m_P^3}\,.
\end{equation}

Eq. (\ref{nB/s-1}) expresses the baryon
asymmetry in terms of parameters characterizing the nonlinear
electrodynamics. In the standard case, i.e. $\gamma=0$, $\eta_B$ vanishes
and no net baryon asymmetry can be generated, as expected.

Introducing the dimensionless parameter $\Gamma\equiv \gamma
[GeV]^{4(1-\delta)}$, Eq. (\ref{nB/s-1}) can be rewritten as
\begin{equation}\label{Gamma}
  \Gamma \left(\frac{B}{GeV^2}\right)^{2\delta} = N\eta_B
  \frac{GeV}{T_D} \left(\frac{M_*}{GeV}\right)^2
  \left(\frac{m_P}{GeV}\right)^3\,,
\end{equation}
where
 \[
N\equiv \sqrt{\frac{\pi g_*}{5}}\frac{2^\delta}{8g_b
(\beta_{RD}-2)\delta (\delta-1)}\,.
 \]
The bound $\eta_B \lesssim 9 \times 10^{-11}$ and Eq.
(\ref{Gamma}) give a constraint (upper bound) on the free
parameter $\gamma$ for fixed magnetic field strengths. For our
estimates, we use the following values of parameters: As pointed
out in \cite{steinhardt}, a possible choice of the cutoff scale
$M_*$ is $M_*=m_{Pl}/\sqrt{8\pi}$ if $T_D=M_I$, where $M_I\sim 2
\,\,10^{16}GeV$ is the upper bound on the tensor mode fluctuation
constraints in inflationary scale \cite{riotto1}. For $T_D$, we
use the decoupling temperature at the GUT scale, $T_D\sim
10^{16}GeV$ (a decoupling temperature at the GUT scale is
phenomenologically acceptable if the unwanted relics like
gravitinos decoupled at the Planck scale so that they will be
diluted away during inflation and will not be regenerated at
reheating at the GUT scale). 

\begin{figure}
\resizebox{7cm}{!}{\includegraphics{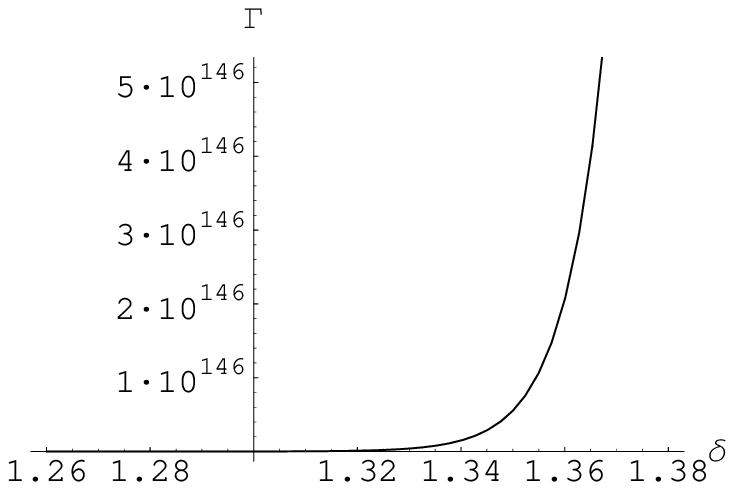}}
\caption{$\Gamma$ vs $\delta$. $\Gamma$ is plotted for
$\beta_{RD}-2=-\frac{1}{2}-2\delta-
\sqrt{4\delta^2-2\delta-\frac{7}{4}}$ and $B=10^{-10}$G.}
 \label{Gamma1}
\end{figure}
\begin{figure}
\resizebox{7cm}{!}{\includegraphics{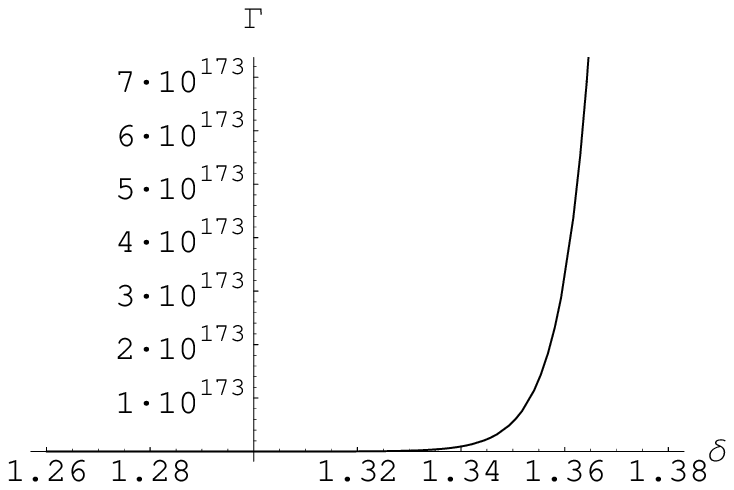}}
\caption{$\Gamma$ vs $\delta$. $\Gamma$ is plotted for
$\beta_{RD}-2=-\frac{1}{2}-2\delta-
\sqrt{4\delta^2-2\delta-\frac{7}{4}}$ and $B=10^{-20}$G.}
 \label{Gamma1a}
\end{figure}
\begin{figure}
\resizebox{7cm}{!}{\includegraphics{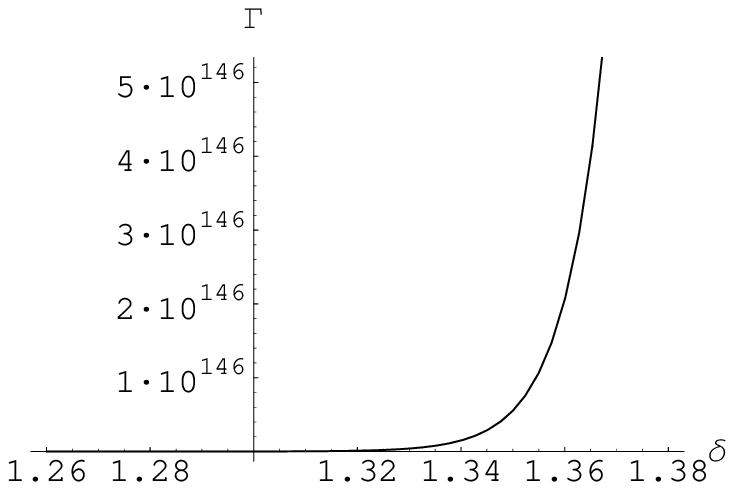}}
\caption{$\Gamma$ vs $\delta$. $\Gamma$ is plotted for
$\beta_{RD}-2=-\frac{1}{2}-2\delta+
\sqrt{4\delta^2-2\delta-\frac{7}{4}}$ and $B=10^{-10}$G.}
 \label{Gamma2}
\end{figure}
\begin{figure}
\resizebox{7cm}{!}{\includegraphics{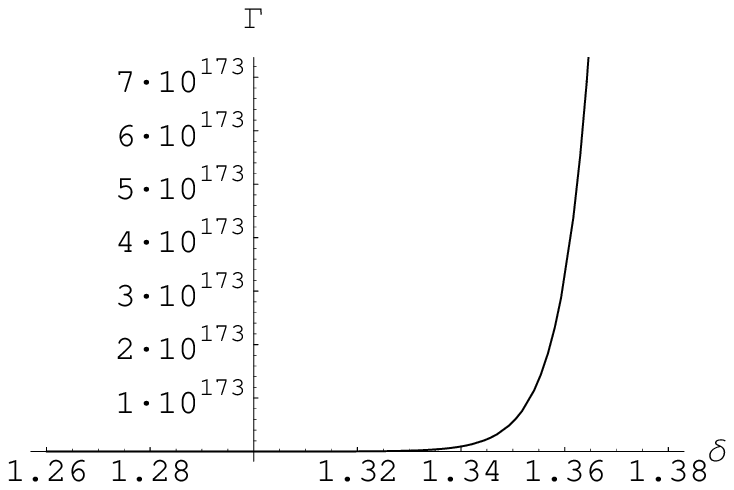}}
\caption{$\Gamma$ vs $\delta$. $\Gamma$ is plotted for
$\beta_{RD}-2=-\frac{1}{2}-2\delta+
\sqrt{4\delta^2-2\delta-\frac{7}{4}}$ and $B=10^{-20}$G.}
 \label{Gamma2a}
\end{figure}

In Figs. \ref{Gamma1} and
\ref{Gamma1a}, and Figs. \ref{Gamma2} and \ref{Gamma2a} are
represented the behavior of $\Gamma$ vs $\delta$ for magnetic
field strength $B\sim 10^{-20}G$ and $B\sim 10^{-10}G$. Figs.
\ref{Gamma1}, \ref{Gamma1a} correspond to $\beta_{RD}=\beta_1$,
whereas Figs. \ref{Gamma2}, \ref{Gamma2a} correspond to
$\beta_{RD}=\beta_2$. The range for $\delta$ is $[1.28, 1.38]$
derived in previous Section (see Tables I and II).

As a final comment, we analyze the validity of our approximation
$\rho_B < \rho_{rad}$. In the regime we worked, $\gamma
X^{\delta-1}\gg 1$, see Section II, the energy density of the
electromagnetic field reads $\rho_B\sim \gamma (B^2/2)^\delta$. By
making use of Eq. (\ref{Gamma}) and $\Gamma\equiv \gamma
[GeV]^{4(1-\delta)}$, we get

\begin{equation}\label{B-eta}
  \Gamma \left(\frac{B}{GeV^2}\right)^{2\delta}=N\eta_B
  \left(\frac{M_*}{GeV}\right)^2
  \left(\frac{m_P}{GeV}\right)^3 \frac{GeV}{T_D} \,.
\end{equation}

The condition $\rho_{rad} > \rho_B$ gives the lower bound on the
temperature $T$
\begin{equation}\label{Tbound}
 T > 1.1 \times
 10^{15}\left(\frac{10^{16}GeV}{T_D}\right)^{1/4}GeV
\end{equation}
where we have used $\displaystyle{\frac{30N}{2^\delta \pi^2
g_*}}\sim {\cal O}(10^{-2})$. As Eq. (\ref{Tbound}) shows, our
assumptions are consistent for temperatures of the
Universe varying in the range $T_{RH} > T >  {\tilde T}$, i.e. the
nonlinear electrodynamics effects are active at GUT scales. In
this regime, nonlinear electrodynamics allows to account for both the
amplification of the primordial magnetic fields and the origin of
the baryon asymmetry.

\section{The Novello-Bergliaffa-Salim (NBS) model of 
nonlinear electrodynamics}

In the framework of nonlinear electrodynamics, we shall now analyze
the Novello-Bergliaffa-Salim (NBS) model
\cite{novello}. This model is particularly interesting because the
nonlinear terms of the electromagnetic field give rise to a
"fluid" with an asymptotically negative equation of state.
Therefore, the accelerated expansion of the Universe can be
attributed to these nonlinear corrections to the standard
electromagnetic Lagrangian.

The Lagrangian of the nonlinear electrodynamics of the NBS model
is \cite{notaNBS}
\begin{equation}\label{salim}
  L_\mu =-X -\frac{\mu^8 }{X}\,,
\end{equation}
where $[\mu]=(energy)^2$. It corresponds to $C=1$, $\delta=-1$ and
$\gamma=\mu^8$ in (\ref{Lagr}).

To derive an upper bound on the parameter $\mu$, NBS assume that
dark energy can be described by the nonlinear term, and using the
current value for $\Omega_{de}=\rho_{de}/\rho_{cr}$, where
$\rho_{cr}=3H_0^2/8\pi G$ is the critical energy density, they
find \cite{novello}
\begin{equation}\label{mu}
  \mu^4\lesssim 3.74 \times 10^{-28}\frac{gr}{cm^3}=1.683
  \times 10^{-45}GeV^4\,.
\end{equation}

The extremely small value of $\mu$ implies a negligible
corrections to Maxwell's electromagnetism. Nonetheless,
one should keep in mind that for extremely low magnetic
field strength is the $1/F$-term of the NBS Lagrangian
that dominates.

\subsection{Primordial Magnetic Field}

In studying the amplification of the magnetic fields, we follow
Section II. In order to obtain the required value $r\sim 10^{-37}$
corresponding to the observed values of the galactic magnetic
field, we assume that the NBS nonlinear electromagnetism is turned
off during the de Sitter era, and turns on at reheating era, till
the time ${\tilde \eta}$ of the radiation-dominated era. The wave
equation for $F$ is given by (\ref{EqWaveFa}) with $\delta = -1$.
As before, we assume that the $(F^2/a^4)$-term is dominant.

\subsubsection{\bf Inflationary de Sitter (dS) phase}

If during this era the nonlinear electrodynamics effects are absent, then the
wave equation for $F$ is $\Box_\eta F= 0$, whose solution is
$F\sim \sin k\eta$ (the solution is independent whether $k\eta
\gtrless 1$ as a consequence of the conformal invariance of the
minimally coupled electromagnetic field \cite{turner}). In the
long wavelength approximation, one obtains

\begin{equation}\label{FdS-NBS}
  F\sim \eta \sim a^{-1} \,.
\end{equation}

\subsubsection{\bf Reheating (RH) phase}

In this phase of the evolution of the Universe,
the wave equation (\ref{EqWaveFa}) admits the solution

\begin{equation}\label{FRH-NBS}
  F \sim  a^{(19\pm \sqrt{105})/4}\,.
\end{equation}

\subsubsection{\bf Radiation-dominated (RD) phase}

During the RD era, finally, the solution for $F$ is given by

\begin{equation}\label{FRD-NBS}
  F \sim  a^{(9\pm \sqrt{17})/2}\,.
\end{equation}

Values of the parameter $r$ are obtained using Eq. (\ref{r}) with the exponents
$p$ and $q$ given by Eqs. (\ref{FdS-NBS}) and (\ref{FRH-NBS}),
$p=-1$ and $q=(19\pm \sqrt{105})/4$. Results are reported in Table
III.

\begin{table}
\caption{Values of $r$ at 1 Mpc and for different ${M_{GUT},
T_{RH}, T_*}$.}
\begin{ruledtabular}
\begin{tabular}{llll}
  $M_{GUT}$(GeV) & $T_{RH}$(GeV) & $T_*$(GeV) & $r$  \vspace{0.09in} \\ \hline\hline
  $10^{17}$ & $10^{15}$ & $10^{15}$ & $10^{-38}$  \\
            & $10^{16}$ & $10^{15.5}$ & $10^{-37}$ \\
            & $10^{16}$ & $10^{16}$ & $10^{-47}$ \\
            & $10^{17}$ & $10^{16}$ & $10^{-37}$ \vspace{0.09in}\\ \hline
  $10^{17}$ & $10^{9}$ & $10^{12}$ & $10^{-42}$ \\
  $10^{16}$ & $10^{9}$ & $10^{12}$ & $10^{-53}$ \\
\end{tabular}
\end{ruledtabular}
\end{table}

It is interesting to notice that the NBS model allows for an amplification
of the magnetic fields. In particular, we can see that the required amplification
(leading to $r\sim 10^{-37}$) may occur for the set of values
 \[
\{M_{GUT}, T_{RH}, T_*\}=\{(10^{17}, 10^{15}, 10^{15}),
 \]
 \[
(10^{17}, 10^{16}, 10^{15.5}), (10^{17}, 10^{17}, 10^{16})\}GeV\,.
 \]

\subsection{Baryon Asymmetry}

Let us now investigate the baryon asymmetry in the framework of
NBS model.

The trace of the energy-momentum tensor for the NBS nonlinear
electrodynamics Lagrangian (\ref{salim}) reads
\begin{equation}\label{traceNBS}
  T^{(NBS)}=\rho-3p=\frac{8\mu^8}{X}\,,
\end{equation}
which is obtained by averaging the magnetic (and electric) field
on a sufficiently large time-dependent three volume
 \begin{equation}\label{average}
 \overline{E}_i=0\,,\quad \overline{B}_i=0\,,\quad \overline{E_i
 B_j}=0\,,
 \end{equation}
 \[
 \overline{E_iE_j}=-\frac{{\bf E}^2}{3}\delta_{ij}\,,\quad
 \overline{B_iB_j}=-\frac{{\bf B}^2}{3}\delta_{ij}\,.
 \]
As in the case discussed in Section III, we assume the background
evolves as in the standard Cosmology, which means that the energy
density of the magnetic field is lesser that the energy density of
radiation. The time
derivative of the Ricci scalar is given by
\begin{equation}\label{derivateR}
  {\dot R}=-\frac{128(5\pm \sqrt{17})}{2}\frac{\mu^8}{B^2}\frac{H}{m_P^2}\,,
\end{equation}
where $H={\dot a}/{a}$. By using again the Einstein field equations (\ref{einstein}), the
net baryon asymmetry generated by nonlinear electrodynamics turns
out to be
\begin{equation}\label{etaNovello}
  \eta_B = N' \frac{\mu^8}{B^2}\frac{T_D}{M_*^2 m_P^3}\,,
\end{equation}
where $N'=N|_{\delta=-1}$. $\eta_B$ vanishes as $\mu=0$. The
observed baryon asymmetry can be generated provided that the
temperature at which the NBS nonlinear electrodynamics is active
satisfies the constraint (\ref{Tbound}), that is at GUT scales.

If we consider $\mu$ as a free parameter, which does not satisfy
Eq. (\ref{mu}), then bounds on $\mu$ from (\ref{etaNovello})
follows by using the previous values of the parameters $M_*\sim
10^{16}GeV$, $T_D \sim 10^{16}GeV$, and a fixed magnetic field
strength. For example, for $B\sim 10^{-20}G$, one obtains $\mu^4
\lesssim 10^{-12}GeV^4$. On the other hand, if one assumes that
the bound (\ref{mu}) holds for conformal time $\eta$ such that
$\eta_{RD}< \eta < {\tilde \eta}$, then to obtain the observed
baryon asymmetry the magnetic field strength must be of the order
$\gtrsim 10^{-54}G$, which seems to be no cosmologically
interesting.

\section{Conclusion}

In this paper we have studied the amplification of the magnetic
field and the origin of the baryon asymmetry in the framework of
the nonlinear electrodynamics. In particular we have analyzed
Lagrangian densities of the form ${\cal L}\sim X+\gamma X^\delta$
and ${\cal L}\sim X+\mu^8/X$. The baryon asymmetry is generated
by means of the (gravitational) coupling between baryon current and
curvature of the background, which is non vanishing during the
radiation dominated era owing the nonlinear effects in the
electromagnetism.

For the Lagrangian of the form $X+\gamma X^\delta$, which we have
studied in the regime in which the nonlinear term dominates the
standard $X$-term, and for the de Sitter, reheating and
radiation dominated eras, we have found that the amplification of
the primordial magnetic field occurs provided that the parameter
$\delta$ falls in the range $[1.26; 1.38]$. Moreover, the analysis
has been performed also for the origin of the baryon asymmetry
occurring during the radiation dominated era.

As concerns the model proposed by Novello-Bergliaffa-Salim, with ${\cal L}\sim X+\mu^8/X$,
the analysis of the amplification of the primordial magnetic fields shows that the required
values $r\sim 10^{-37}$, necessary for explaining the observed galactic magnetic fields, is obtained provided that the electromagnetic nonlinear terms turn on at the reheating era, but are zero at the de Sitter epoch.

In conclusion, the nonlinear electrodynamics, which is the reduction in the Abelian sector of an effective model of the low energy (3+1)-QCD \cite{pagel}, seems a promising candidate for
studying cosmological scenarios which go beyond the standard cosmology and particle physics.

\acknowledgments H. J. M. C. thanks FAPERJ, Brazil for financial
support and ICRANet Coordinating Centre, Pescara, Italy for
hospitality during the early stages of this work. G.L. acknowledges
the financial support of MIUR through PRIN 2006 Prot.
$1006023491_{-}003$, of a contract with the Agenzia Spaziale
Italiana, and of research funds provided by the University of
Salerno.


\begin{thebibliography}{99}
\bibitem{notaY} The more general expression for $L$ includes a dependence
                on $Y$, i.e. $L=L(X, Y)$, where $Y=\frac{1}{4}F_{\mu\nu} {}^*
                F^{\mu\nu}$, ${}^*F^{\mu\nu}=\frac{1}{2\sqrt{-g}}
                \varepsilon^{\mu\nu\alpha\beta}F_{\alpha\beta}$ ($\epsilon_{0123}=1$).
                In this paper we set $Y=0$.
\bibitem{noteJ} $J^\mu$ may be replaced by any current which yields a net $B-L$ charge in
        thermal equilibrium; here $B$ and $L$ stand for baryon and lepton number, respectively.
\end{thebibliography}
\end{document}